\documentclass[%
 preprint, 
 pre,
 amsmath,amssymb,
 aps, physrev,
]{revtex4-2}

\usepackage{graphicx}
\usepackage{dcolumn}
\usepackage{bm}
\usepackage{xcolor}
\usepackage[normalem]{ulem} 
\usepackage{cancel}         

\begin{document}

\preprint{}

\title{\textbf{Suppression of the Laser Hosing Instability in a Parabolic Plasma Channel}}

\author{Yash M Jogiya}
 \affiliation{School of Advanced Sciences, Vellore Institute of Technology, Vellore-632014, Tamil Nadu India}
 \author{Vishwa Bandhu Pathak}%
 \email{Contact author: vishwa.bandhu@vit.ac.in}
\affiliation{%
School of Advanced Sciences, Vellore Institute of Technology, Vellore-632014, Tamil Nadu India}%


\begin{abstract}
We present an analytical stability analysis of the laser hosing instability in a parabolic channel in the weakly relativistic regime. We employ a variational principle approach to derive a set of coupled equations describing laser dynamics in a transversely inhomogeneous plasma. The quadratic Lagrangian formalism is applied to perform a stability analysis and to derive the growth rates for hosing in both long- and short-wavelength regimes. We find that a parabolic channel suppresses the long-wavelength hosing mode completely, for any channel depth, and that the unstable band of the short-wavelength mode is bounded below by a cutoff wavenumber set by the ratio of the channel frequency to the hosing coupling. A steeper channel narrows the unstable band and lowers the short-wavelength growth rate, but cannot stabilize the mode, since the dispersion relation forbids damping identically. The analysis is accompanied by two-dimensional particle-in-cell simulations, which reproduce the prediction in the nonlinear regime: the long-wavelength mode is absent in a channel, the surviving modulation shortens as the channel steepens, and its amplitude falls with channel depth. Further, our simulation results demonstrate the sensitivity of off-axis laser propagation to the plasma channel and distinguish the hosing instability from laser oscillations due to off-axis laser propagation in a plasma channel.

\end{abstract}

\maketitle


\section{\label{intro}Introduction}

The nonlinear interaction of a high-intensity short-pulse laser as it propagates through an underdense plasma is an active area of research. As an intense pulse propagates, it can experience a spatio-temporal evolution of its field structure. Understanding this process is crucial due to its relevance to various applications such as laser wakefield acceleration (LWFA) \cite{Tajima_1979, Esarey_2009, Lu2007, Pathak_2012, Pathak_2018}, radiation generation schemes \cite{Hojbota_2023}, and laser-driven fusion \cite{Tabak_1994, ICF_persp}. Work in these areas has identified numerous self-modulated processes, such as Raman forward scattering (RFS) instabilities \cite{RFS_mori}, relativistic self-focusing \cite{SF_Gupta}, envelope self-modulation \cite{SM_esarey}, and hosing \cite{duda_hosing_1999, ren2001}. The spatiotemporal growth rates of these instabilities have been established \cite{mori1997, sprangle_1994, shvets1994}. It has been observed that the laser evolution through a few Rayleigh lengths of plasma follows a common sequence. As the faster-growing instabilities saturate, the final nonlinear state of the laser is dominated by the long-wavelength hosing (LWH) instability \cite{duda_hosing_1999}. Moreover, laser hosing instability can be predominant in regimes employing longer laser pulses, such as inertial confinement fusion \cite{ICF_persp}, laboratory astrophysics \cite{solar_burst}, and self-modulated LWFA \cite{SMLWFA}. Recent experiments \cite{exper_evid} have confirmed the dominance of long-wavelength hosing for a few-femtosecond laser pulse propagating in the plasma, revealing a density-dependent scaling of the hosing wavelengths.

The hosing instability persists at near-critical plasma densities and higher laser intensities where an ultra-intense laser pulse is used; correspondingly, several methods have been proposed to suppress and control the hosing process. It is proposed that the hosing instability can be mitigated by using the multiple pulse method or by the use of ultra-relativistic pulses \cite{Ceurvorst_2016}. In the direct laser acceleration (DLA), the rapidly growing hosing instability leads to a random tilt in electron beams, which can be suppressed by employing the parabolic plasma channel \cite{Huang_2017}. However, in the case of a weakly relativistic laser pulse over a few Rayleigh lengths of propagation, the transverse asymmetries in laser intensity or the plasma density cause a tilt in the laser trajectory. The front of the laser oscillates transversely about the original laser axis at nearly twice the plasma wavelength, while at a position further back, the laser modulates at longer wavelengths \cite{duda_hosing_1999}. This hosing grows with propagation and in turn degrades the pointing stability and controllability of the laser pulse. Furthermore, when the laser intensity is low for self-guiding, transverse plasma inhomogeneities are often introduced to guide the laser pulse over the acceleration distance far beyond the Rayleigh length \cite{ehrlich1996, sprangle_1992}. These transverse density profiles may significantly change the laser dynamics.

The primary goals of this paper are to investigate the effects of plasma channel profiles on the growth of laser hosing instability and to distinguish between modulation of the laser pulse due to plasma density asymmetry and the hosing instability. We follow the variational principle approach developed by Duda \textit{et al} \cite{duda_2000}, which parametrizes hosing directly and exhibits the coupling between the centroids of the laser and the plasma wake as the underlying mechanism \cite{duda_2000, duda_hosing_1999}, together with its extension to a transversely inhomogeneous plasma by Pathak \textit{et al} \cite{Pathak_2018} and \cite{yash2026}, from which we take the trial functions and density integrals $I_{ij}$ used in Sec.~II. In a uniform plasma, the centroid pair carries no restoring term. For an ultra-intense pulse in a preformed channel, Huang \textit{et al} \cite{Huang_2017} obtain a centroid equation in which the channel enters as a restoring force, but do not solve it for the growth rate. We carry the formulation into a stability analysis of the parabolic channel, obtaining closed-form growth rates in both wavelength regimes, establishing that the long-wavelength mode cannot exist at any channel depth, and fixing the cutoff wavenumber that bounds the surviving band. The analytical results are accompanied by particle-in-cell (PIC) simulations using OSIRIS \cite{fonseca2002, Fonseca_2008}.

This paper is organized as follows: In Section II, a brief introduction to the variational model and a set of coupled differential equations are presented. In Section III, the perturbation expansion of the Lagrangian is performed for the parabolic plasma channel case to analyze the hosing instability. In Section IV, the growth rate for hosing and the cutoff condition are derived. In addition, PIC simulations using OSIRIS \cite{fonseca2002, Fonseca_2008} are presented to examine laser hosing in the parabolic channel. Section V presents conclusions.

\section{\label{variation}Variational principle approach}

The variational principle method reformulates a system of partial differential equations into an action integral\textemdash{Hamiltonian principle}, $S = \int d^4 \vec{x} \text{ }\mathcal{L}$, where $\mathcal{L}$ is the Lagrangian density of the system, and $\vec{x} \equiv (\vec{r}, t)$. The action is stationary for first-order variations of the dependent variables: the laser vector potential $\vec{a}$ and the plasma scalar potential $\phi$. An approximate model of the system is then constructed by introducing trial functions of $\vec{a}$ and $\phi$ defined by descriptive parameters that depend on the coordinates of the co-moving frame $(\psi = ct-z, \tau=z)$. Then we normalize all the time dimensions to $\omega_p^{-1}$ and the space dimensions to $k_p^{-1} \equiv c/\omega_p$, where $c$ is the speed of light in vacuum, $\omega_p=\sqrt{n_0e^2/(m\epsilon_0)}$ is the plasma frequency for a plasma of density $n_0$ with $e$ and $m$ as the electron charge and rest mass, and $\epsilon_0$ as the permittivity of free space. The next step involves integrating over the transverse coordinates to obtain a Lagrangian that depends solely on the independent variables $(\psi, \tau)$, making the trial function parameters new dependent variables. Further, varying the action or solving the Euler-Lagrange equation for this reduced Lagrangian with respect to the new dependent variables provides a set of approximate, coupled differential equations for the descriptive parameters. Detailed derivations of the method are given in \cite{duda_2000, Pathak_2018}. Here we start with the approximate trial functions for $\vec{a}$ and $\phi$ characterized by a set of descriptive parameters \cite{duda_2000, Pathak_2018},

\begin{equation}
\begin{split}
    a =& \ A(\psi, \tau)exp\left[ik_x(\psi, \tau)(x - x_a(\psi,\tau)) + ik_y(\psi, \tau)(y-y_a(\psi, \tau))\right] \\&
     \times exp\left[-\left\{1-i\alpha_x(\psi, \tau)\right\}\frac{{x-x_a(\psi, \tau)}^2}{W_{xa}(\psi, \tau)^2}\right]
     \times exp\left[-\left\{1-i\alpha_y(\psi, \tau)\right\}\frac{{y-y_a(\psi, \tau)}^2}{W_{ya}(\psi, \tau)^2}\right] ,
    \label{trial_fn_a}
\end{split}
\end{equation}

\begin{align}
\begin{split}
    \phi=\Phi(\psi, \tau) \times exp\left[-2\frac{{x-x_\phi(\psi, \tau)}^2}{W_{x\phi}(\psi, \tau)^2}
    -2\frac{{y-y_\phi(\psi, \tau)}^2}{W_{y\phi}(\psi, \tau)^2}\right] ,
    \label{trial_fn_phi}
\end{split}
\end{align}

where, $A$ is a complex amplitude as $A(\psi, \tau)=\sqrt{\xi(\psi,\tau)} exp[i\chi(\psi,\tau)]$. $x_a(\psi,\tau)$, $y_a(\psi,\tau)$ and $x_\phi(\psi,\tau)$, $y_\phi(\psi,\tau)$ are the transverse centroid coordinates in the x-y direction for $a$ and $\phi$, respectively. $\alpha_x(\psi,\tau)$, $\alpha_y(\psi,\tau)$ are the curvatures of the laser wave-front, and $k_x(\psi,\tau)$, $k_y(\psi,\tau)$ are the transverse laser wavenumbers. The spot sizes are represented by $W_{xa}(\psi,\tau)$, $W_{ya}(\psi,\tau)$, $W_{x\phi}(\psi,\tau)$, $W_{y\phi}(\psi,\tau)$  for $a$ and $\phi$. The trial function for $\phi$ is chosen to ensure that, for a non-evolving wakefield, $\partial_\psi \phi = 0$, the scalar potential reduces to the ponderomotive potential, $|\phi| = |a|^2/4$ \cite{duda_2000}.

Substituting the trial functions from equations \eqref{trial_fn_a} and \eqref{trial_fn_phi} into the Lagrangian \cite{Pathak_2018},

\begin{equation}
    \mathcal{L} =\nabla_\bot a\cdot\nabla_\bot a^\ast - ik_0\left(a\partial_\tau a^\ast - a^\ast \partial_\tau a\right) -2\left(\partial_\psi \phi\right)^2+2\phi^2+\left[f(\vec{r_\bot})-\phi\right]aa^\ast ,
\end{equation}

where $f(\vec{r_\bot})$ is a transverse inhomogeneity function, and integrating the resulting expression over the transverse coordinates yields the new Lagrangian \cite{Pathak_2018}. This new Lagrangian is defined in terms of the laser and plasma parameters, such as $A$, $\Phi$, $\alpha_x$, $ k_x$, $x_a$, $W_{xa}$, etc., forming a new set of variables which are functions of $\psi$ and $\tau$. We solve the Euler-Lagrange equation \cite{goldstein2001},

\begin{equation}
    \frac{\partial\mathcal{L}}{\partial X}-\frac{\partial}{\partial\psi}\left(\frac{\partial\mathcal{L}}{\partial(\partial_\psi X)}\right)-\frac{\partial}{\partial\tau}\left(\frac{\partial\mathcal{L}}{\partial(\partial_\tau X)}\right) = 0
    \label{eul_lag_eq}
\end{equation}

for the parameters $X \equiv \chi$, $\alpha_x$, $\alpha_y$, $k_x$, $k_y$ gives a set of simplified equations \cite{duda_2000, Pathak_2018}. These include an equation for laser power conservation \cite{duda_2000},

\begin{equation}
    \partial_\tau P = 0,
    \label{power}
\end{equation}

Equation \eqref{power} gives $P=A_0^2 W_0^2$, where $A_0$ and $W_0$ are the initial normalized vector potential amplitude and normalized spot size of the laser. Throughout, $P_{cr} \simeq 17\,(\omega_0/\omega_p)^2$~GW denotes the critical power for relativistic self-focusing, so that in these normalized units $P/P_{cr} = P/32$. For wavefront curvatures defined as the evolution of the spot size,
\begin{equation}
    \alpha_x = -\frac{k_0}{4}\partial_\tau W_{xa}^2,\text{ } \alpha_y = -\frac{k_0}{4}\partial_\tau W_{ya}^2,
    \label{curvature}
\end{equation}

For the transverse wavenumbers in terms of the laser centroid motion,
\begin{equation}
    k_x = -k_0\partial_\tau x_a, \text{ } k_y = -k_0\partial_\tau y_a,
    \label{wavenumber}
\end{equation}

By substituting these derived equations \eqref{power}-\eqref{wavenumber} into the Lagrangian, the variables $\alpha_x$, $\alpha_y$, $k_x$, and $k_y$ can be eliminated. Leaving out the ignorable variable $\chi$, the laser power $P$ can be treated explicitly as a constant. This leads to the form of the Lagrangian mentioned in equation \eqref{lagrangian} \cite{Pathak_2018}.

\begin{equation}
    \langle{\mathcal{L}}\rangle = \langle{\mathcal{L}}\rangle_0 + \frac{P}{2} \frac{I_{00} [x_a, W_{xa}, W_{ya}]}{W_{xa}W_{ya}},
    \label{lagrangian}
\end{equation}

where, $\langle{\mathcal{L}}\rangle_0$ is the Lagrangian obtained by Duda \cite{duda_2000} as shown below,

\begin{align}
\begin{split}
    \langle{\mathcal{L}}\rangle_0 =&\text{ } \frac{P}{2} \left( \frac{1}{W_{xa}^2} +\frac{1}{W_{ya}^2} \right) -\frac{P k_0^2}{8}[\left( \partial_\tau W_{xa} \right)^2
    +\left( \partial_\tau W_{ya} \right)^2+4\left( \partial_\tau x_{a} \right)^2+4\left( \partial_\tau y_{a} \right)^2]
     +\frac{W_{x\phi}W_{y\phi}\Phi^2}{2}  \\&
    -\frac{W_{x\phi} W_{y\phi} \Phi P}{2\sqrt{\left(W_{xa}^2+W_{x\phi}^2 \right) \left(W_{ya}^2+W_{y\phi}^2\right)}}
    \times exp \left[-2 \biggl\{\frac{\left(x_a-x_\phi\right)^2}{W_{xa}^2+W_{x\phi}^2}+\frac{\left(y_a-y_\phi\right)^2}{W_{ya}^2+W_{y\phi}^2}\biggr\}\right] \\&
    -\frac{W_{x\phi}W_{y\phi}}{2}\left(\partial_\psi\Phi\right)^2
    -\frac{3}{8}\Phi^2\biggl\{\frac{W_{y\phi}}{W_{x\phi}}(\partial_\psi W_{x\phi})^2 + \frac{W_{x\phi}}{W_{y\phi}}(\partial_\psi W_{y\phi})^2\biggr\} \\&
    -\Phi^2 \biggl\{\frac{W_{y\phi}}{W_{x\phi}}(\partial_\psi x_{\phi})^2+ \frac{W_{x\phi}}{W_{y\phi}}(\partial_\psi y_{\phi})^2\biggr\}
    -\frac{\partial_\psi(\Phi^2)}{4}\partial_\psi\left(W_{x\phi}W_{y\phi}\right)-\frac{\Phi^2}{4}\partial_\psi W_{x\phi}\partial_\psi W_{y\phi},
\label{reduced lag}
\end{split}
\end{align}

and the integral $I_{ij}$ is defined as \cite{Pathak_2018},

    \begin{align}
    \begin{split}
        I_{ij} =& \frac{2}{\pi} \int \int \left[f(x, y)-1\right](x-x_a)^i (y-y_a)^j
        \times exp\left[-2\frac{(x-x_a)^2}{W_{xa}^2}-2\frac{(y-y_a)^2}{W_{ya}^2}\right]dx dy
    \label{integral}
    \end{split}
    \end{align}

where, $i$ and $j$ are integers and $f(x,y)$ is transverse inhomogeneity. $I_{00}$ in equation \eqref{lagrangian} explicitly represents the contribution of plasma inhomogeneity in the laser-plasma interaction, with $i=j=0$. To obtain the final set of differential equations, the Euler-Lagrange equation \eqref{eul_lag_eq} is solved for the remaining parameters. For further analytical simplicity in the model, identical variations in laser-plasma spot sizes are considered, i.e., $W_{xa} = W_{x\phi}$ and $W_{ya} = W_{y\phi}$. By solving the Euler-Lagrange equations for the centroids $x_a$, $y_a$, $x_\phi$, $y_\phi$, and spot sizes $W_{xa}$ and $W_{ya}$, the following equations governing the evolution of the respective parameters are derived.

\begin{align}
\begin{split}
    \partial^2_\tau x_a + \frac{2}{k^2_0 W_{xa}^3 W_{ya}} \left[I_{10} +\frac{P}{16}(x_a-x_\phi)\right] = 0
    \label{laser cent}
\end{split}
\end{align}

\begin{align}
\begin{split}
    (\partial^2_\psi +1)x_\phi - x_a = 0
    \label{plasma cent}
\end{split}
\end{align}

\begin{align}
\begin{split}
    \partial^2_\tau W_{xa} - \frac{4}{k^2_0 W_{xa}^3} \Biggl[1 -\frac{P}{32}\frac{W_{xa}}{W_{ya}} \biggl\{1-\frac{3(x_a-x_\phi)^2}{W_{xa}^2}
    -\frac{(y_a-y_\phi)^2}{W_{ya}^2}+\frac{64I_{20}}{PW_{xa}^2}\biggr\}
    +\frac{I_{00}}{2}\frac{W_{xa}}{W_{ya}} \Biggr] = 0
    \label{x spotsize}
\end{split}
\end{align}

\begin{align}
\begin{split}
    \partial^2_\tau W_{ya} - \frac{4}{k^2_0 W_{ya}^3} \Biggl[1 -\frac{P}{32}\frac{W_{ya}}{W_{xa}} \biggl\{1-\frac{(x_a-x_\phi)^2}{W_{xa}^2}
    -\frac{3(y_a-y_\phi)^2}{W_{ya}^2}+\frac{64I_{02}}{PW_{ya}^2}\biggr\}
    +\frac{I_{00}}{2}\frac{W_{ya}}{W_{xa}} \Biggr] = 0
    \label{y spotsize}
\end{split}
\end{align}

The equations for $y_a$ and $y_\phi$ are identical to those for $x_a$ and $x_\phi$, respectively, but independent of each other. However, the spot size equations are coupled and signify that the unidirectional transverse inhomogeneities in the plasma can influence the laser self-focusing along the plasma inhomogeneity and in the orthogonal direction as well. The centroid equations are consistent with \cite{Pathak_2018}. The extension of this section lies in Eqs. \eqref{x spotsize} and \eqref{y spotsize}, which retain a general inhomogeneity in both transverse directions through $I_{ij}$, rather than a gradient along $x$ alone. Both centroid offsets are consequently kept in both equations, where they enter with unequal weights according to whether the offset is parallel or orthogonal to the spot size in question. The resulting spot-size evolution can be symmetric or asymmetric based on the inhomogeneity introduced in those directions. In the absence of any plasma inhomogeneities, the spot-size becomes $W_{xa} = W_{ya} = W_a$, and equations \eqref{x spotsize} and \eqref{y spotsize} reduce to the well-known self-focusing equation \cite{mori1997}, with the threshold power condition as $P/P_{cr} = A^2W_a^2/32$ \cite{Esarey_2009, anderson1979}.

The set of coupled differential equations \eqref{laser cent}-\eqref{y spotsize} derived in this section provides a complete, nonlinear description for the evolution of the laser spot-sizes and centroids in a transversely inhomogeneous plasma. However, obtaining a full analytical solution to this complex system is generally intractable. Therefore, to gain physical insight into the stability of the laser propagation against small perturbations, we now shift our focus from the full nonlinear dynamics to a linear stability analysis.

\section{\label{sec:quadratic}Quadratic Lagrangian Approach}

In order to investigate the linear stability properties of the system with transversely inhomogeneous plasma, we derive the quadratic Lagrangian by performing a perturbative expansion about a known zeroth-order solution, with the underlying methodology remaining the same as \cite{duda_2000}. We consider that the zeroth-order solution is slowly varying compared to the time scale of the perturbation growth, allowing us to neglect its derivatives in the linearized formulation. The Lagrangian is expanded in powers of a small perturbation parameter $\delta$. Solving the Euler-Lagrange equations for the first-order terms merely reproduces the zeroth-order equations of motion; thus, these terms can be omitted. The second-order terms, which are quadratic in the first-order perturbation, yield a variational principle that governs the linearized dynamics. This quadratic Lagrangian encapsulates the essential features of the linearized system and provides a convenient and systematic approach to derive the corresponding stability equations. The primary advantage of this approach is that applying the Euler-Lagrange equations to the quadratic Lagrangian directly yields a set of linearized equations of motion for the perturbations. This simplifies the analysis and also provides clear physical insight by naturally decoupling the system into its fundamental normal modes of instability.

For analytical simplicity in the transversely inhomogeneous plasma case, the density integral is first solved for the linear transverse density gradient \cite{Pathak_2018} with the profile defined by $f(x) = 1 + N_c(x - x_{c0})$, where at $x = x_{c0}$ the plasma density is $n_0$. When the plasma profile is a simple linear density gradient, contributions due to plasma inhomogeneity enter the variational formulation through linear coefficients. Indeed, for this profile $I_{00}/(W_{xa}W_{ya}) = N_c(x_a - x_{c0})$, which is independent of the spot sizes, so no new spot-size coupling can arise. All terms that survive after expanding the action to the second order in the perturbation parameter $\delta$ are still quadratic in the first-order perturbations, but introduce no new couplings. Consequently, the quadratic Lagrangian obtained is identical to the uniform density case presented in \cite{duda_2000}, and the parametric equations remain the same, following the same growth rate of the instabilities.

In contrast, when the parabolic guiding channel is introduced, the density profile is given by
\begin{align}
    f(x, y) = 1 + \frac{1}{W_{ch}^2}[(x - x_{c0})^2 + (y - y_{c0})^2]
    \label{parabolic profile}
\end{align}

where $W_{ch}$ is the density scale radius of the parabolic channel. $x_{c0},$ $y_{c0}$ are the transverse channel-axis coordinates in the x and y directions, respectively, at time $\tau=0$. The plasma channel introduces density curvature, so the equilibrium itself contains second-order derivatives. When expanding the action, the parabolic profile generates additional $\delta^2$ contributions. It is worth mentioning that the zeroth-order solution depends on the on-axis and off-axis initial positions of the laser. However, the channel contribution in the quadratic Lagrangian is independent of the initial laser position relative to the channel axis. This follows directly from Eq.~\eqref{integral}: for the profile of Eq.~\eqref{parabolic profile},
\begin{align}
    I_{00} = \frac{W_{xa}W_{ya}}{W_{ch}^{2}}\left[(x_a-x_{c0})^{2}+(y_a-y_{c0})^{2}+\frac{W_{xa}^{2}+W_{ya}^{2}}{4}\right],
    \label{I00 channel}
\end{align}
and on substituting $x_a = \delta x_{a1}$ the offsets $x_{c0}$, $y_{c0}$ appear only at orders $\delta^{0}$ and $\delta^{1}$. They therefore cancel out of the quadratic Lagrangian entirely, and the hosing growth rate cannot depend on where the laser is launched relative to the channel axis. This is the formal basis for the off-axis result presented in Sec.~\ref{sec:hosing} and for the fact that the $\delta^2$ terms survive only because the profile is curved. These surviving $\delta^{2}$ terms carry the restoring force of the channel into the stability analysis.

Now to obtain the quadratic Lagrangian, we use the perturbation substitutions \cite{duda_2000},

\begin{equation}
\label{perturbations}
\begin{gathered}
    \Phi = \frac{P}{4W_0^2} + \delta\Phi_1 \\[2ex]
    \begin{aligned}
        W_{xa} &= W_0 + \delta W_{xa1}, &\qquad\qquad W_{ya} &= W_0 + \delta W_{ya1}, \\
        W_{x\phi} &= W_0 + \delta W_{x\phi1}, &\qquad\qquad W_{y\phi} &= W_0 + \delta W_{y\phi1}, \\[1ex]
        x_a &= \delta x_{a1}, &\qquad\qquad y_a &= \delta y_{a1}, \\
        x_\phi &= \delta x_{\phi1}, &\qquad\qquad y_\phi &= \delta y_{\phi1}.
    \end{aligned}
\end{gathered}
\end{equation}

Then, Taylor expansion of the Lagrangian to second order in $\delta$ and collecting the second-order terms:

\begin{equation}
\begin{aligned}
    \mathcal{L}_{\delta^{2}} = &- \frac{16}{W_{0}^{4}}\frac{P}{P_{cr}} \bigg[\biggr\{k_{0}^{2}W_0^4[(\partial_\tau x_{a1})^2 + (\partial_\tau y_{a1})^2] + \frac{4P}{P_{cr}}[(\partial_\psi x_{\phi1})^2 + (\partial_\psi y_{\phi1})^2] \\
    &- \frac{4P}{P_{cr}}[(x_{a1} - x_{\phi1})^2 + (y_{a1} - y_{\phi1})^2]\bigg\}\biggr]
    - \frac{W_{0}^{2}}{2}\{(\partial_{\psi} \Phi_1)^{2} - \Phi_{1}^{2}\} \\
    &- \frac{4}{W_{0}}\frac{P}{P_{cr}}\{(\partial_\psi W_{x\phi1} + \partial_\psi W_{y\phi1})\partial_\psi \Phi_1 - (W_{xa1}+ W_{ya1} + W_{x\phi1} + W_{y\phi1})\Phi_1\} \\
    &- \frac{4}{W_{0}^{4}}\frac{P}{P_{cr}}\bigg\{k_{0}^{2}W_0^4[(\partial_\tau W_{xa1})^2 + (\partial_\tau W_{ya1})^2] - 2(6 - P/P_{cr}) (W_{xa1}^2 + W_{ya1}^2) + 4\frac{P}{P_{cr}}W_{xa1} W_{ya1}\bigg\} \\
    &- \frac{8 }{W_{0}^{4}}\bigg(\frac{P}{P_{cr}}\bigg)^2[3\{(\partial_\psi W_{x\phi1})^2 + (\partial_\psi W_{y\phi1})^2 - W_{x\phi1}^2 - W_{y\phi1}^2\} \\
    &+ 2\{(\partial_\psi W_{x\phi1}) (\partial_\psi W_{y\phi1}) - W_{x\phi1}W_{y\phi1} \}]
    - \frac{16}{W_0^2}\bigg(\frac{P}{P_{cr}}\bigg)^2(W_{xa1}-W_{ya1})(W_{x\phi1}-W_{y\phi1}) \\
    &+ \frac{4}{W_{ch}^2}\frac{P}{P_{cr}} \{W_{xa1}^2 +W_{ya1}^2 + 4(x_{a1}^2+y_{a1}^2) \}
    \label{quad lag}
\end{aligned}
\end{equation}

Now, we define the variables as,

\begin{align}
    \overline{w_a} \equiv \frac{1}{2}(W_{xa1} + W_{ya1}), \overline{w_\phi} \equiv \frac{1}{2}(W_{x\phi1} + W_{y\phi1})
    \label{avg spot}
\end{align}

\begin{align}
    \Delta w_{a} \equiv \frac{1}{2}(W_{xa1} - W_{ya1}),  \Delta{w_\phi} \equiv \frac{1}{2}(W_{x\phi1} - W_{y\phi1})
    \label{diff spot}
\end{align}

Then, the quadratic Lagrangian can be written as three separate sub-Lagrangians as

\begin{align}
    \mathcal{L} =  \mathcal{L}_{hos} + \mathcal{L}_{sym-smod} + \mathcal{L}_{asym-smod}
    \label{sub lag}
\end{align}

where,

\begin{equation}
\begin{aligned}
    \mathcal{L}_{hos} = &- \frac{64}{W_{0}^{4}}\frac{P}{P_{cr}}\bigg[r_L^2\{(\partial_\tau x_{a1})^2 + (\partial_\tau y_{a1})^2
    - \omega_{ch}^2(x_{a1}^2+y_{a1}^2) \} \\
    &+ \frac{P}{P_{cr}}\{(\partial_\psi x_{\phi1})^2 + (\partial_\psi y_{\phi1})^2\}
    - \frac{P}{P_{cr}}\{(x_{a1} - x_{\phi1})^2 + (y_{a1} - y_{\phi1})^2\}\bigg]
    \label{hosing lag}
\end{aligned}
\end{equation}

\begin{equation}
\begin{aligned}
    \mathcal{L}_{sym-smod} = &- \frac{W_{0}^{2}}{2}\{(\partial_{\psi} \Phi_1)^{2} - \Phi_{1}^{2}\}
    - \frac{8}{W_{0}}\frac{P}{P_{cr}}\{(\partial_\psi \overline{w_\phi})\partial_\psi \Phi_1 - (\overline{w_a}+ \overline{w_\phi})\Phi_1\} \\
    &- \frac{32}{W_0^4}\frac{P}{P_{cr}}\bigg[r_L^2\{(\partial_\tau \overline{w_a})^2 - \omega_{ch}^2\overline{w_a}^2 \}
    - \biggr(3 - \frac{P}{P_{cr}}\biggr) \overline{w_a}^2 + 2\frac{P}{P_{cr}}\{(\partial_\psi \overline{w_\phi})^2 - \overline{w_\phi}^2 \}\bigg]
    \label{ssm lag}
\end{aligned}
\end{equation}

\begin{equation}
\begin{aligned}
    \mathcal{L}_{asym-smod} = &- \frac{32}{W_{0}^{4}}\frac{P}{P_{cr}}\bigg[r_L^2 \{(\partial_\tau \Delta w_a)^2 - \omega_{ch}^2\Delta w_a^2\} -3 \Delta w_a^2 \\
    &+ \frac{P}{P_{cr}} \{(\partial_\psi \Delta w_\phi)^2 - \Delta w_\phi^2 + 2\Delta w_\phi \Delta w_a \}\bigg]
    \label{assm lag}
\end{aligned}
\end{equation}

Here, $r_L = k_0W_0^2/2$ is the normalized Rayleigh length of the laser pulse and $\omega_{ch} = 1/(k_0W_{ch})$ is the normalized channel frequency of the plasma channel.

The quadratic Lagrangian separates into three independent subsystems corresponding to three distinct normal modes. The first subsystem $\mathcal{L}_{hos}$ has contributions only from the centroid parameters $x_a$ and $x_\phi$ and is relevant for the hosing instability. The second subsystem $\mathcal{L}_{sym-smod}$ involves the evolution of the averaged spot sizes and $\Phi$, characterizing the symmetric self-modulation. The third subsystem $\mathcal{L}_{asym-smod}$ is for the antisymmetric differentials between the spot sizes, which describe the antisymmetric self-modulation. Although a full analysis of all three modes is possible, we will examine the hosing instability for the parabolic plasma channel exclusively; the two self-modulation branches in a channel are deferred to a separate study, since they involve the coupling to $\Phi_1$ and do not share the clean two-oscillator structure that makes the hosing problem tractable in closed form.

\begin{figure}[!t]
    \centering
    \includegraphics[width=1.0\linewidth]{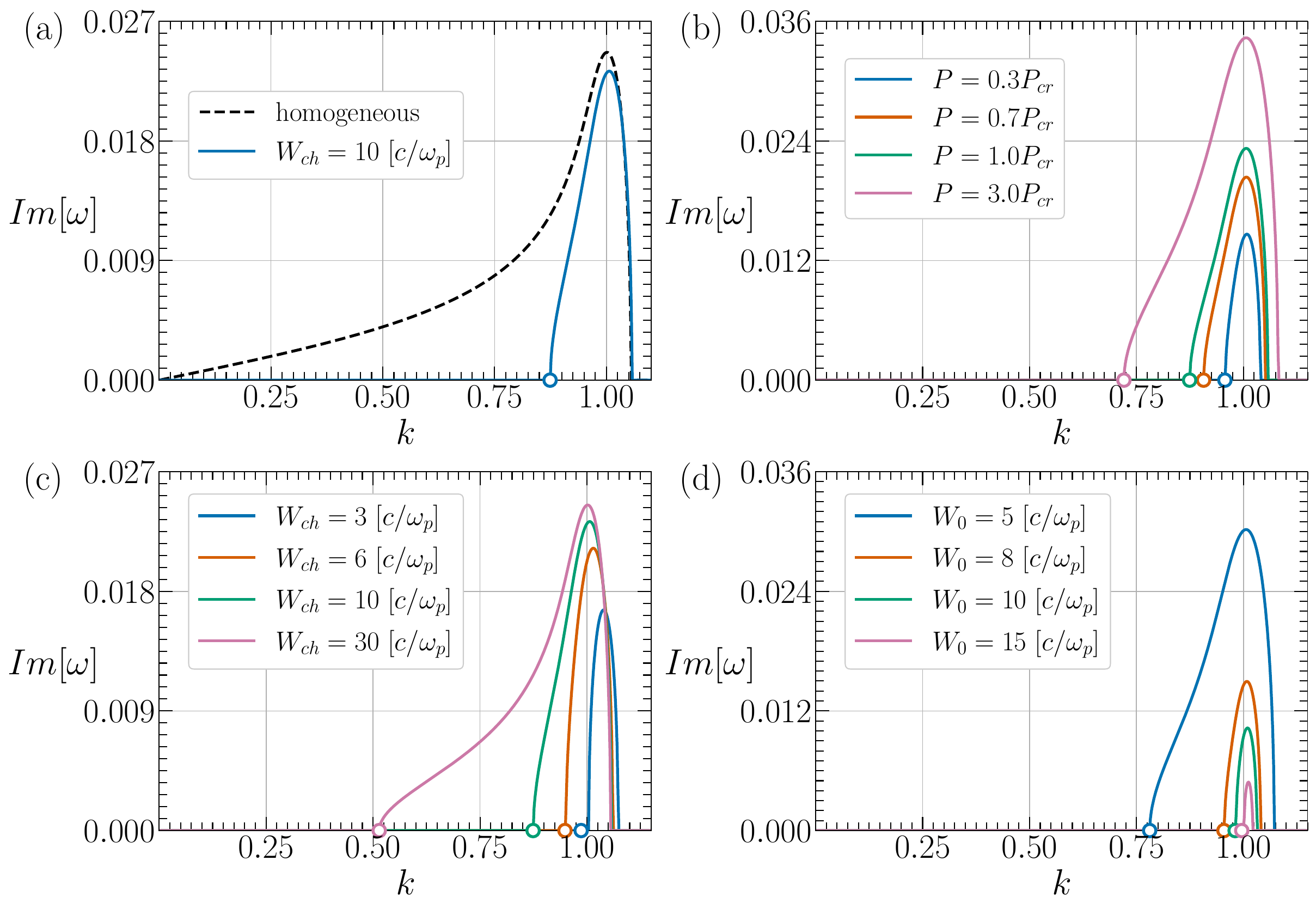}
    \caption{Hosing growth rate against wave number. (a) Homogeneous plasma (dashed line) compared with a parabolic channel (solid line). (b) Parabolic channel at normalized powers $P/P_{cr}=0.3$, $0.7$, $1.0$ and $3.0$. (c) Channel radii $W_{ch}=3c/\omega_p$, $6c/\omega_p$, $10c/\omega_p$ and $30c/\omega_p$. (d) Spot sizes $W_0=5c/\omega_p$, $8c/\omega_p$, $10c/\omega_p$ and $15c/\omega_p$. Except where varied, $W_0=6c/\omega_p$, $W_{ch}=10c/\omega_p$ and $P/P_{cr}=1$. The marker on each curve is the cutoff wave number $k_{cut}$ below which the channel is stable. The laser frequency is fixed at $\omega_0=8\omega_p$.}
    \label{fig:hosing growth}
\end{figure}

\section{\label{sec:hosing}Hosing Instability}

We start by solving the Euler-Lagrange equation for $x_{a1}$ and $x_{\phi1}$ using $\mathcal{L}_{hos}$, which yields the linearized hosing equations:

\begin{equation}
    \partial_\tau^2 x_{a1} + (\kappa^2 + \omega_{ch}^2) x_{a1} = \kappa^2 x_{\phi1}
    \label{hos eq1}
\end{equation}

\begin{equation}
    \partial_\psi^2 x_{\phi1} + x_{\phi1} = x_{a1}
    \label{hos eq2}
\end{equation}

where, $\kappa=\sqrt{P/(P_{cr} r_{L}^2)}=A_0/(2\sqrt{2}k_0 W_0)$. Equations~\eqref{hos eq1} and \eqref{hos eq2} are a pair of coupled oscillators: the laser centroid is restored at frequency $\sqrt{\kappa^{2}+\omega_{ch}^{2}}$ and the wake centroid at the plasma frequency, and the instability is the resonance between them. The channel enters only through the restoring term of Eq.~\eqref{hos eq1}; it stiffens the laser centroid without adding any drive, and that asymmetry is the mechanism of the hosing suppression. For $y_{a1}$ and $y_{\phi1}$ the equations are identical. In the absence of the plasma channel, the above equations reduce to the known equations for hosing \cite{sprangle_1994, shvets1994}. To evaluate the growth rate and the range of unstable wavelengths for hosing, we first transform the system into laboratory coordinates $(t, z)$ by using the transformation $\partial_\tau \rightarrow{\partial_t + \partial_z}$ and $\partial_\psi \rightarrow{\partial_t}$, where $(t, z)$ are normalized by $(1/\omega_p, c/\omega_p)$. Assuming solutions to equations \eqref{hos eq1} and \eqref{hos eq2} of the form $exp[i(kz - \omega t)]$, we obtain the dispersion relation,

\begin{equation}
(\omega^2-1)(\omega-k)^2 = \omega^2 \biggr(\kappa^2+\omega_{ch}^2 \biggr) - \omega_{ch}^2 ,
\label{dis rel}
\end{equation}

where, $\omega$ and $k$ of the hosing mode are normalized with $\omega_p^{-1}$ and $c/\omega_p$ respectively. Equation \eqref{dis rel} confirms that the coupling of the plasma wave and laser mode drives hosing instability. Written out, Eq.~\eqref{dis rel} is the quartic

\begin{align}
  \omega^{4}-2k\,\omega^{3}+\left(k^{2}-1-\kappa^{2}-\omega_{ch}^{2}\right)\omega^{2}
  +2k\,\omega+\left(\omega_{ch}^{2}-k^{2}\right)=0 ,
  \label{quartic}
\end{align}

whose coefficients are real for real $k$. Its roots are therefore real or form complex-conjugate pairs. Hence $\max\mathrm{Im}[\omega]$ can never be negative, and hosing can at best be brought to marginal stability, never damped. First, we use this dispersion relation to examine the growth rate of the system, and solve for the complex $\omega$ as a function of real $k$. Figure \ref{fig:hosing growth}(a) shows a comparison of the hosing growth rate from this solution in a uniform plasma \cite{duda_hosing_1999} (dashed line) and a parabolic channel (solid line) with $W_{ch} = 10 c/\omega_p$ for a matched laser beam with $P/P_{cr} = 1$, $\omega_0 = 8 \omega_p$, $W_0 = 6c/\omega_p$ using the above equation \eqref{dis rel}.
The low-$k$ tail present in the uniform case is absent in the channel, so that long-wavelength hosing does not occur there. Near the peak at $k \sim 1$, by contrast, the two curves nearly coincide. The channel therefore does not reduce the growth rate uniformly. It removes the band of wavenumber and leaves the short-wavelength peak nearly intact. The influence of the laser power within a plasma channel is detailed in Figure \ref{fig:hosing growth}(b). As the normalized power $P/P_{cr}$ increases from 0.3 to 3.0 for the channel with $W_{ch}=10 c/\omega_p$, both the peak growth rate and the range of unstable wave numbers increase substantially. Physically, hosing is a feedback loop between the two centroids. A transversely displaced pulse drives a displaced wake, and the refractive-index gradient of that wake deflects the pulse further in the same direction. Power enters Eq. \eqref{hos eq2} only through the coupling $\kappa$, which measures the strength of this loop and grows with the intensity of the pulse, whereas the restoring term $\omega_{ch}^{2}$ supplied by the channel is independent of it. Increasing the power thus strengthens the destabilizing half of the competition alone, and widens the unstable band. Complementing this, figure \ref{fig:hosing growth}(c) illustrates the critical role of the channel radius. It should be noted that $W_{ch}$ does not represent the physical radius of the channel, but the characteristic transverse scale over which the density increases. For a fixed power of $P/P_{cr}=1.0$, the growth rate is shown for the channel radius $W_{ch} = 3c/\omega_p$, $6c/\omega_p$, $10c/\omega_p$, $30c/\omega_p$. This figure shows that a steeper plasma channel density profile leads to a lower growth rate and a shorter range of unstable wave numbers. However, the sharp density variation drives faster oscillations in the laser centroid, shifting the growth rate peak slightly towards higher wave numbers. As the channel becomes wider, the guiding effect weakens as the laser beam interacts with a lower density variation. In this case, the growth rate tends to approach that of the homogeneous plasma. Finally, Figure \ref{fig:hosing growth}(d) shows the growth rate for several values of the laser spot sizes $W_0=5c/\omega_p$,$ 8c/\omega_p$, $10c/\omega_p$, $15c/\omega_p$ with $W_{ch}=10c/\omega_p$ for matched power $P/P_{cr}=1$. When the spot size is comparable to or larger than the channel radius, the hosing is largely suppressed, occurring only under resonance condition.

\begin{figure}[!t]
    \centering
    \includegraphics[width=1.0\linewidth]{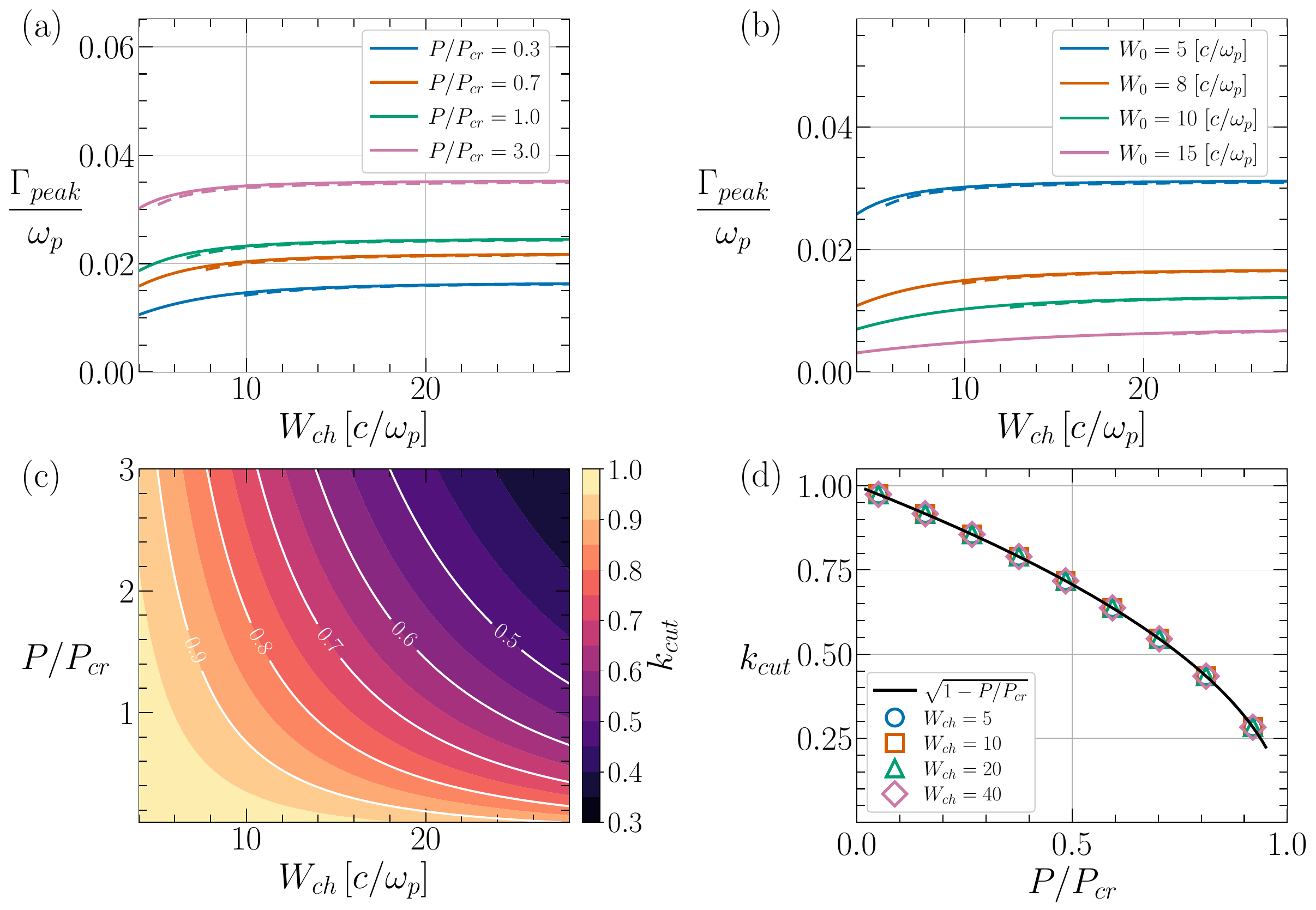}
    \caption{Peak hosing growth rate under the resonance condition as a function of the channel radius, (a) for normalized laser powers $P/P_{cr} = 0.3$, $0.7$, $1.0$, and $3.0$ at $W_0 = 6c/\omega_p$, and (b) for laser spot sizes $W_0 = 5c/\omega_p$, $8c/\omega_p$, $10c/\omega_p$, and $15c/\omega_p$ at the matched condition $P/P_{cr} = 1$. In (a) and (b), the solid curves are the exact peak of $\mathrm{Im}[\omega]$ obtained from Eq. \eqref{dis rel}, and the asymptotic form Eq. \eqref{peak growth} is drawn dashed over the valid range of $\mathcal{V} < 0.5$. (c) presents Cutoff wavenumber $k_{cut}$ over the span of channel radius and power at fixed $W_0 = 6c/\omega_p$. (d) The same quantity at the matched spot size $W_m$, for $W_{ch}=5$, $10$, $20$ and $40\,c/\omega_p$. Matching cancels the dependence on channel radius, and the four curves collapse onto Eq.~\eqref{kcut matched}, drawn solid. The laser frequency is fixed at $\omega_0=8\omega_p$.}
    \label{fig:res_cut_lwh}
\end{figure}

To derive the peak-growth-rate $(\Gamma_{peak})$ when $Re[\omega]\sim k \sim 1 $, we consider $\omega = \omega_r + \epsilon$ , where $\omega_r$ is the real part of $\omega$ and $\epsilon$ is small complex shift such that $|\epsilon| \ll \omega_r$, and $Im[\epsilon]=\Gamma_{peak}$. Thus, from the equation \eqref{dis rel}, under the condition $Re[\omega]\sim k \sim 1 $ we obtain the cubic $2\epsilon^{3} - 2\epsilon(\kappa^{2}+\omega_{ch}^{2}) - \kappa^{2} = 0$, whose root of positive imaginary part gives

\begin{equation}
\begin{aligned}
    \Gamma_{peak} = \frac{\sqrt{3}}{2} \biggr[\biggr(\frac{\kappa^2}{2}\biggr)^{1/3} -\frac{1}{3}{\big(\kappa^2 + \omega_{ch}^2\big)}{\biggr(\frac{2}{\kappa^2}\biggr)^{1/3}}\biggr] .
    \label{peak growth}
\end{aligned}
\end{equation}

Equation~\eqref{peak growth} follows from a single Newton step about the leading-order root $2\epsilon_0^{3} = \kappa^{2}$, and so holds only while 

\begin{align}
  \mathcal{V} \;\equiv\; \frac{\kappa^{2}+\omega_{ch}^{2}}{\left(\kappa^{2}/2\right)^{2/3}} \;\ll\; 1 .
  \label{validity}
\end{align}

Since $\kappa^{2} \propto W_0^{-4}$ while $\omega_{ch}^{2}$ grows as the channel tightens, $\mathcal{V}$ becomes large for a large spot in a narrow channel. Where it does, the expansion fails, and Eq.~\eqref{peak growth} can return a negative growth rate, which Eq.~\eqref{quartic} forbids. In Fig.~\ref{fig:res_cut_lwh} it is therefore drawn only for $\mathcal{V}<0.5$, alongside the exact peak of $\mathrm{Im}[\omega]$ obtained from Eq.~\eqref{dis rel}.

Figures \ref{fig:res_cut_lwh}(a) and (b) both show the peak growth rate against the channel radius, (a) scanning the normalized power at fixed $W_0 = 6c/\omega_p$, and (b) the laser spot size at matched power, and the same asymptotic behavior observed in the dependence of the growth rate on the channel radius which can be understood analytically by examining its partial derivative, $\partial\Gamma_{peak}/\partial W_{ch} = [\sqrt{3}k_0^2W_{ch}^3(\kappa^2/2)^{1/3}]^{-1}$. As this gradient is inversely proportional to $W_{ch}^3$, the peak growth rate initially increases rapidly, and the slope vanishes for large channel radius, indicating strong saturation. How far the saturation goes depends on the parameters, with the reduction being largest where $\kappa$ is smallest, so that the channel lowers the peak by the greatest factor at low power and for large spot sizes, which are the cases in which the hosing is weakest. In both panels, a tighter channel reduces the peak growth rate monotonically and narrows the unstable band. The reduction saturates, however, and the growth rate never changes sign, as Eq. \eqref{quartic} requires. The parameter range in which Eq. \eqref{peak growth} would return a negative value is precisely that in which $\mathcal{V} \gg 1$ and the expansion no longer holds. Furthermore, there exists a cutoff for the hosing to occur, which can be obtained by substituting $\omega=k$ in equation \eqref{dis rel}, tracking the unperturbed laser mode when the hosing growth rate $Im[\omega] = 0$,

\begin{equation}
    k_{cut} = \sqrt{\frac{\omega_{ch}^2}{\omega_{ch}^2 + \kappa^2}}
    \label{cutoff}
\end{equation}

The cutoff condition provides a means to restrict the number of growing hosing modes by tuning the channel radius. The variation of the cutoff wavenumber from equation \eqref{cutoff} is shown in Figure \ref{fig:res_cut_lwh}(c), as a function of the normalized power and the channel radius. This shows that even shallow channels with relatively small density variations can efficiently suppress the long-wavelength hosing. Furthermore, as the channel profile steepens, the bandwidth of the unstable hosing modes progressively narrows, restricting it to about $k \sim 1$. However, complete suppression of short-wavelength hosing (SWH) remains unattainable. Together, these plots show that the mitigation of hosing instability requires precise tuning of the plasma channel radius with the laser spot size and power.

For a matched beam, the cutoff of Eq. \eqref{cutoff} reduces to a function of the power alone. The spot size that a parabolic channel guides without envelope oscillation follows from the vanishing of the bracket in Eq.~\eqref{x spotsize} as $W_m = \sqrt{2W_{ch}}\,(1-P/P_{cr})^{1/4}$. This differs from the familiar matching rule $W_0 = W_{ch}$, which applies when $W_{ch}$ is a physical channel radius rather than, as here, the scale length over which the density varies. Imposing this matched condition in Eq.~\eqref{cutoff}, the explicit dependence on $W_0$ and $k_0$ cancels and the cutoff is fixed by the normalized power alone,

\begin{align}
  k_{cut}\Big|_{\rm matched} = \left[1 + \frac{P/P_{cr}}{1-P/P_{cr}}\right]^{-1/2}
  = \sqrt{1-P/P_{cr}} \, .
  \label{kcut matched}
\end{align}

Figure \ref{fig:res_cut_lwh}(d) shows the cutoff evaluated for four channel radii falling onto the single curve of Eq. \eqref{kcut matched}. A matched beam well below the critical power therefore has almost no unstable band, while a broad band of growing modes requires operation close to $P/P_{cr}=1$, where relativistic self-focusing competes with the channel. This sets a practical limit on how far hosing can be suppressed by channel guiding alone while retaining a matched beam.

Next, analytical evidence for the suppression of long-wavelength hosing is obtained by evaluating the growth rate for the long-wavelength regime. Taking the approximation $|\omega| \ll 1$, solving equations \eqref{hos eq1} and \eqref{hos eq2} gives the growth rate in the long-wavelength regime,

\begin{equation}
    \Gamma_{LWH} = \frac{\sqrt{\kappa^2k^2-\omega_{ch}^2(1 +\kappa^2)}}{1+\kappa^2}
    \label{long growth}
\end{equation}

The onset of LWH instability dictates that the term under the square root must be positive. However, in the long-wavelength regime, $k \sim 0$, and the condition always remains $k^2\kappa^2 < \omega_{ch}^2 (1+\kappa^2)$. Consequently, the real part of the growth rate vanishes, $Re[\Gamma_{LWH}] =0$, indicating that the presence of a preformed plasma channel entirely suppresses the development of the LWH instability and stable laser propagation can be achieved.

This holds even without the long-wavelength approximation. Writing $\Gamma = i(k - \omega)$ for the growth rate in $\tau$, so that $\mathrm{Re}[\Gamma] = \mathrm{Im}[\omega]$, Eq. \eqref{dis rel} becomes

\begin{align}
  \Gamma^{2}(\omega) = \frac{\kappa^{2}\omega^{2}}{1-\omega^{2}} - \omega_{ch}^{2} ,
  \label{rigid shift}
\end{align}

in which the first term is the homogeneous result, and the channel enters only as the constant $-\omega_{ch}^2$. As $\omega \to 0$, this tends to $-\omega_{ch}^2$, negative for any channel, so no long-wavelength mode can grow at any channel depth. The cutoff $k_{cut}$ is where $\Gamma^2$ changes sign.

\begin{figure}[b!]
    \centering
    \includegraphics[width=0.5\linewidth]{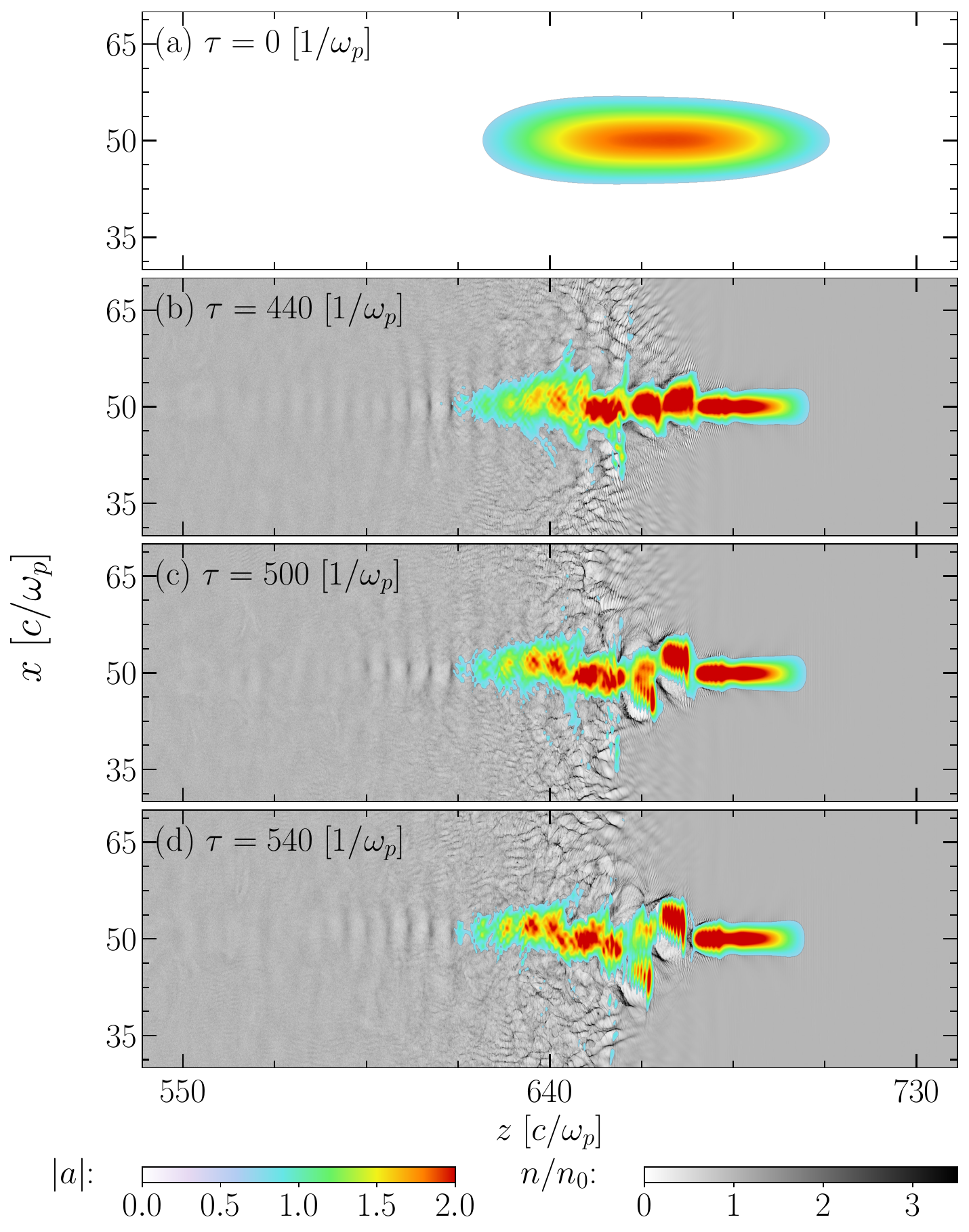}
    \caption{ 2D simulation snapshot of the laser envelope $|a|$ and plasma density in uniform plasma, at (a) $\tau[\omega_p^{-1}]=0$, (b) $440$, (c) $500$ and (d) $540$. Laser parameters are $a_0=2$, $\omega_0=8\omega_p$, envelope length $L=150c/\omega_p$, spot size $W_0=6c/\omega_p$ and initial transverse centroid position $x_{a0}=50c/\omega_p$.}
    \label{fig:unisim}
\end{figure}

To examine the analytical results, we performed 2D PIC simulations using OSIRIS \cite{fonseca2002, Fonseca_2008}. These simulations investigate the development of hosing in the final nonlinear state of short-pulse lasers after propagation over several Rayleigh lengths. The simulations use a moving window frame that advances at c. The simulation box measured $200 \times 100 (c/\omega_p)^2$ in the longitudinal and transverse directions, on an $8250 \times 1600$ z-x Cartesian grid with $16$ particles per cell. The ions are modelled as an immobile neutralizing background. The laser is initialized with a transverse Gaussian profile and a polynomial in the longitudinal direction. The peak normalized vector potential is $a_0 = 2$, the laser frequency $\omega_0 = 8.0\omega_p$, the pulse envelope length $L = 150c/\omega_p$ with the corresponding FWHM $\simeq 75c/\omega_p$, and the spot size $W_0 = 6 c/\omega_p$, focused at the edge of the plasma, which starts at $200 c/\omega_p$ in the z direction. The initial transverse centroid position $x_{a0} = 50c/\omega_p$ is at the central axis of the plasma. Unless otherwise specified, the simulation parameters remain the same for all the results presented in this work.

Figure \ref{fig:unisim} shows a sequence of plots of the laser vector potential, each using a common colour scale. The snapshots correspond to times $\tau [\omega_p^{-1}] = 0$, $440$, $500$ and $540$ in a uniform plasma. The analytical model is valid only under the ideal conditions of the weakly relativistic regime and matched beams. However, since our primary interest is in the nonlinear final stage of the laser evolution, we therefore conduct simulations at a moderately relativistic intensity of $a_0 = 2$. The runs are accordingly outside the formal domain of the linear theory, and are presented as a demonstration of the predicted phenomenology rather than as a quantitative test of Eq.~\eqref{dis rel}. 
Figure \ref{fig:unisim}(a)-(d) follows the pulse into the final state. After a few Rayleigh lengths, the laser self-focuses strongly and the long-wavelength mode comes to dominate. Additionally, Raman forward scattering (RFS) \cite{mori1997} leads to the breakup of the laser envelope into a sequence of beamlet structures. As the laser evolves, the hosing instability amplifies, ultimately disrupting the structure of the laser envelope. This sequence is the one described by Duda \textit{et al} \cite{duda_hosing_1999}, in which self-focusing and RFS first divide the pulse into beamlets spaced at $\lambda_p$ and hosing then displaces successive beamlets in opposite directions, producing a modulation at $2\lambda_p$ that lengthens toward the back of the pulse.

\begin{figure}[t!]
    \centering
    \includegraphics[width=1.0\textwidth]{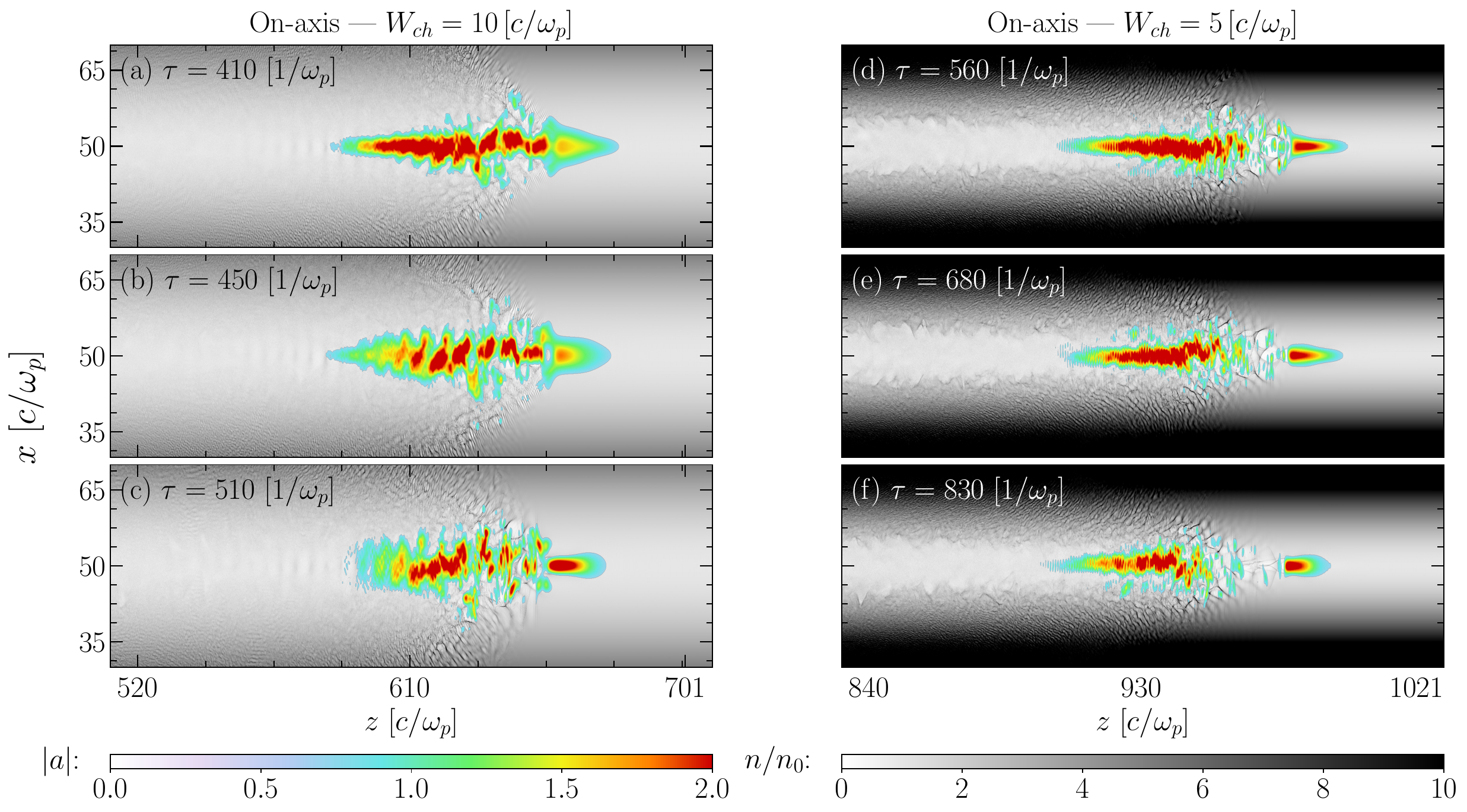}
    \caption{Snapshot of laser and plasma density for two on-axis runs in a parabolic channel. (a) $\tau[\omega_p^{-1}]=410$, (b) $450$ and (c) $510$ for $W_{ch}=10c/\omega_p$, the weaker channel, in which short-wavelength hosing grows. (d) $\tau[\omega_p^{-1}]=560$, (e) $680$ and (f) $830$ for $W_{ch}=5c/\omega_p$, the stronger channel, in which the long-wavelength mode is absent, and the short-wavelength mode is reduced in amplitude. Laser parameters are $a_0=2$, $\omega_0=8\omega_p$, $L=150c/\omega_p$ and $W_0=6c/\omega_p$.}
    \label{fig:parasim}
\end{figure}

We further performed PIC simulations to examine the influence of the parabolic channel on hosing, keeping the laser parameters identical to those used in Figure \ref{fig:unisim}. Figures \ref{fig:parasim}(a)-(c) display snapshots of the laser vector potential in a parabolic plasma with channel radius $W_{ch} = 10c/\omega_p$, where the laser is initialized along the channel axis. The built-in transverse inhomogeneity of the channel removes the long-wavelength hosing in Figure \ref{fig:parasim}, as predicted by Figure \ref{fig:hosing growth}(a). For this wider channel, the short-wavelength peak is reduced only slightly from the uniform-plasma case, so short-wavelength hosing is still clearly present. In contrast, figures \ref{fig:parasim}(d)-(f) present the case of a narrower channel with $W_{ch} = 5c/\omega_p$, which is smaller than the laser spot size $W_0 = 6c/\omega_p$. In this case, after a few Rayleigh lengths of propagation, the front of the laser pulse is depleted due to Raman scattering, whereas the trailing part of the pulse experiences strong self-focusing and channel-focusing. Although signs of hosing are visible, the oscillation amplitude is noticeably reduced compared to the wider-channel case, in agreement with Figure \ref{fig:hosing growth}(c), and with the peak growth rate of Eq. \eqref{peak growth}, both of which fall with channel depth.

\begin{figure}[t!]
    \centering
    \includegraphics[width=1.0\textwidth]{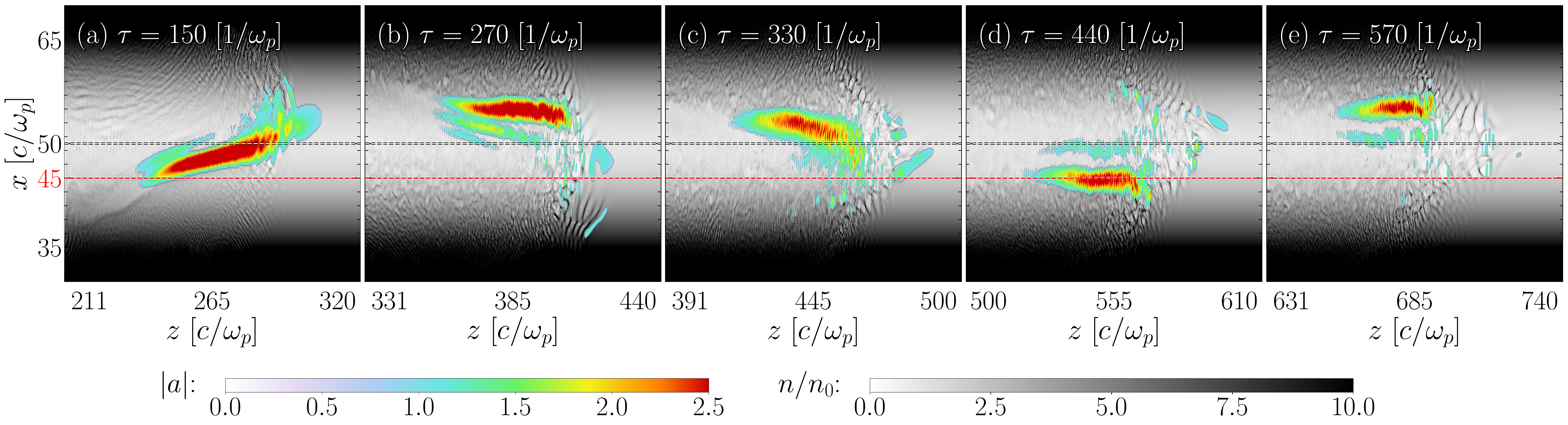}
    \caption{2D simulation snapshot of the laser envelope and the plasma density of off-axis propagation, with the laser launched at $\Delta x_0=-5c/\omega_p$ in a channel of $W_{ch}=5c/\omega_p$, at (a) $\tau[\omega_p^{-1}]=140$, (b) $270$, (c) $330$, (d) $440$ and (e) $570$. The white dashed line marks the channel axis and the red dashed line represents the initial laser position. Over this interval, the pulse is displaced as a whole and follows an oscillatory trajectory about the axis, which is to be distinguished from hosing. Other parameters are as in Figure \ref{fig:parasim}.}
    \label{fig:offaxis}
\end{figure}

Next, we consider the laser propagation in a parabolic channel where the pulse is initialized off the channel axis with $\Delta x_0 = -5c/\omega_p$, presented in Figure \ref{fig:offaxis}. The laser and plasma parameters are the same as used in Figure \ref{fig:parasim}(d). Off-axis injection is a scenario frequently encountered in laser-plasma experiments, where a plasma channel is employed to guide the laser over distances much longer than the Rayleigh length. Figures \ref{fig:offaxis}(a)-(e) covers $\tau = 140\omega^{-1}$ to $570\omega^{-1}$. Over this interval, the pulse is displaced from the axis but develops no hosing, as the on-axis run of figure \ref{fig:parasim}(d) also does not at $\tau = 560\omega^{-1}$. The off-axis launch therefore leaves the hosing behaviour unchanged, as predicted in Sec. \ref{sec:quadratic}, that the initial transverse position cannot influence the onset of hosing. Nevertheless, the laser centroid exhibits an oscillatory trajectory, as previously predicted in \cite{Pathak_2018}. This transverse drift of the laser may be mistaken for hosing in experimental diagnostics or simulations. However, unlike the hosing instability, where localized transverse oscillations within the pulse have a wavelength that increases toward its back, this drift represents a smooth translation of the laser centroid driven by the off-axis placement in the channel. The two are separable in practice: the drift is a rigid displacement of the whole pulse oscillating at the channel frequency $\omega_{ch}$, with no dependence on $\psi$, whereas hosing is a $\psi$-dependent modulation whose wavelength lengthens toward the rear. The distinction between these behaviours is essential for correctly interpreting experimental and simulation data on laser hosing.

\section{Conclusion}

We developed an analytical model based on a variational approach to study the laser dynamics in transversely inhomogeneous plasma within the weakly relativistic regime, deriving a full set of coupled equations for the laser and plasma parameters. Using the Quadratic Lagrangian formalism, we extended the stability analysis for a parabolic plasma channel and identified three normal modes of instability: hosing, symmetric, and asymmetric spot-size self-modulation. For the hosing instability, growth rates were obtained in both the long- and short-wavelength regimes, revealing that the long-wavelength mode is suppressed in a parabolic channel for any channel depth, while the short-wavelength mode persists but can be tuned through a steeper channel gradient. The channel supplies a restoring force on the laser centroid but adds nothing that drives it, so its only effect on the growth rate is a constant reduction applied at every wavenumber. Long-wavelength modes cannot overcome it, while short-wavelength modes survive above a cutoff fixed by the channel, with a peak growth rate that falls steadily as the channel is made steeper but never reaches zero. These analytical predictions were accompanied by two-dimensional PIC simulations, which reproduce the predicted phenomenology in the nonlinear regime. In fusion and electron beam generation experiments, where laser stability is crucial, laser hosing can severely degrade experimental outcomes. Moreover, it is observed that the hosing becomes prominent at higher laser power. The preformed plasma channel can improve the stability and control over the laser propagation. Additionally, we discuss the difference between the laser hosing instability and the laser modulation about the channel axis due to misalignment of the laser injection. 
%
%

\begin{acknowledgments}
The authors would like to acknowledge the support from VIT SEED grant funding SG20220070 and from the Department of Science and Technology, Government of India, under the grant number SR/FST/MS-II/2023/139(C)-VIT Vellore.
\end{acknowledgments}

\bibliography{apssamp}

\end{document}